\documentclass[twocolumn,prl,showpacs]{revtex4}

\usepackage{graphicx}
\usepackage{dcolumn}
\usepackage{bm}
\usepackage{amssymb, amsmath}
\usepackage{color}

\begin{document}
\title{Microscopic interpretation of the Dynes formula for the tunneling
  density of states}

\author{Franti\v{s}ek Herman and Richard Hlubina}
\affiliation{Department of Experimental Physics, Comenius
  University, Mlynsk\'{a} Dolina F2, 842 48 Bratislava,
  Slovakia}

\begin{abstract}
Excellent fits of the tunneling density of states in disordered
superconductors can be often achieved making use of the
phenomenological Dynes formula. However, no consistent derivation of
this formula has been available so far. The Dynes formula can be
interpreted by the simplest causal frequency-dependent gap function
$\Delta(\omega)$ with a vanishing gap at the Fermi level. Here we
show, within the coherent potential approximation, that precisely such
gap function describes superconductors with a Lorentzian distribution
of pair-breaking fields and arbitrary potential disorder. We predict
spectral and thermodynamic properties of such superconductors.
\end{abstract}
\maketitle

{\it Introduction.} The tunneling density of states $N(\omega)$ is a
basic characteristics of the single-particle properties of
superconductors. The knowledge of $N(\omega)$ has played a major role
in identification of the pairing mechanism in conventional
superconductors, and with a similar aim $N(\omega)$ is often studied
in modern superconductors \cite{Fischer07}. On the other hand,
$N(\omega)$ is also used as a diagnostic tool enabling to discover the
existence of pair-breaking processes in superconductors and to
quantify their extent \cite{Menard15}. Such studies are important from
the basic physics point of view, for instance in the context of the
still not completely understood superconductor-insulator transitions
\cite{Gantmakher10,Szabo16}, but also from the point of view of
applied physics, since in many electronic applications of
superconductors such pair-breaking processes are to be avoided
\cite{Wang14}.

The presence of pair-breaking processes shows up in the tunneling
experiment as a finite density of states within the ideal
superconducting gap ${\bar \Delta}$. Long ago, a simple
phenomenological formula has been proposed for superconductors with
such processes \cite{Dynes78},
\begin{equation}
N(\omega)=N_0{\rm Re}\left[(\omega+i\Gamma)/
\sqrt{(\omega+i\Gamma)^2-{\bar \Delta}^2}\right],
\label{eq:dynes}
\end{equation}
which is now known as the Dynes formula. The parameter $\Gamma$ in
this formula quantifies the effect of the pair-breaking processes and
$N_0$ is the normal-state density of states at the Fermi level.  

\begin{figure}[t]
\includegraphics[width=7.0cm]{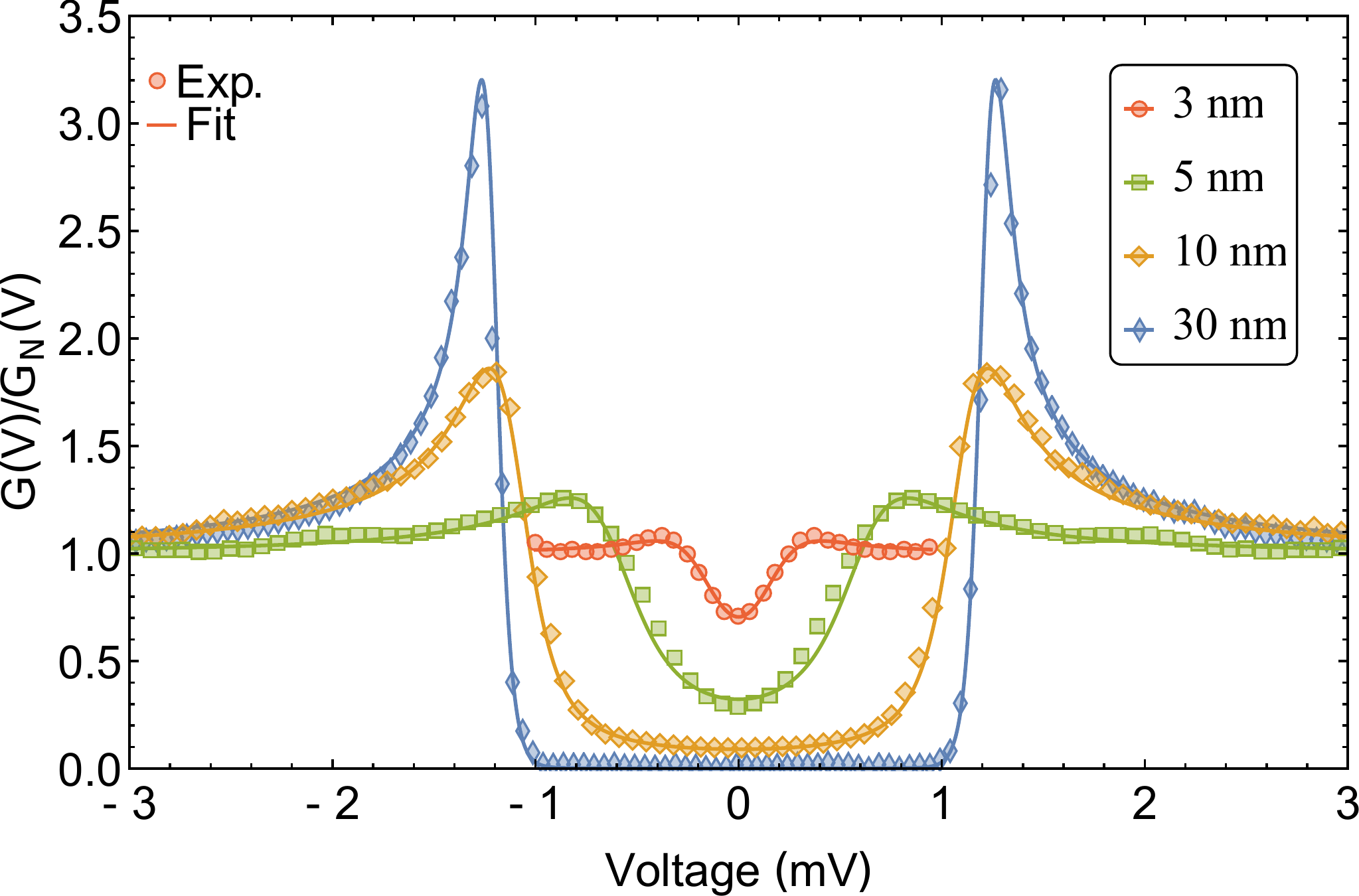}
\caption{(Color online) Normalized tunneling conductance of thin MoC
  films with varying thickness at T$\approx$500~mK, from
  \cite{Szabo16}, with fits to the thermally smeared Dynes
  formula. For further details see \cite{SM}.}
\label{fig:szabo}
\end{figure}

In order to demonstrate the quality of fits which can be achieved
making use of Eq.~(\ref{eq:dynes}), in Fig.~\ref{fig:szabo} we
reproduce the recently measured low-temperature tunneling data on a
series of MoC films with varying thickness \cite{Szabo16}, together
with their fits to the Dynes formula.  Similarly perfect agreement
between experimental data for disordered superconductors and the Dynes
formula has in fact been observed quite frequently, see
e.g. \cite{White86,Sherman15}, indicating that Eq.~(\ref{eq:dynes})
should be caused by a generic mechanism. 

The only mechanism leading to the Dynes formula which has been
suggested so far postulates that its appearance in tunneling
experiments is caused by inelasticity of the tunneling process
\cite{Pekola10}.  However, this mechanism can not explain the
systematic changes of $N(\omega)$ observed in Fig.~\ref{fig:szabo},
which must have a truly intrinsic origin.  The aim of this paper
therefore is to propose a generic and intrinsic microscopic
interpretation of the Dynes formula.

{\it Gap function}. Let us start by noting that, within the Eliashberg
theory, $N(\omega)$ is completely determined by the gap function
$\Delta(\omega)$:
\begin{equation}
N(\omega)=N_0
{\rm Re}\left[\omega/\sqrt{\omega^2-\Delta^2(\omega)}\right].
\label{eq:eliashberg}
\end{equation}
According to Eq.~(\ref{eq:dynes}), $N(\omega)$ is finite at the Fermi
level and this requires that $\Delta(\omega)$ vanishes as
$\omega\rightarrow 0$. The gap function $\Delta(\omega)$ should also
be causal, i.e. analytic in the upper half-plane, and it should
approach ${\bar \Delta}$ in the high-energy limit.  It is known
\cite{Mikhailovsky91} that the simplest function with these
properties,
\begin{equation}
\Delta(\omega)=\omega{\bar \Delta}/(\omega+i\Gamma),
\label{eq:dynes_delta}
\end{equation}
does lead to the Dynes formula, when inserted into
Eq.~(\ref{eq:eliashberg}).  Therefore our task in the rest of this
paper is to find a microscopic explanation of
Eq.~(\ref{eq:dynes_delta}).

It is worth pointing out that Mikhailovsky et
al. \cite{Mikhailovsky91} did find a mechanism leading to
Eq.~(\ref{eq:dynes_delta}). In fact, by a careful analysis of the
Eliashberg equations they have shown that Eq.~(\ref{eq:dynes_delta})
applies even in a clean system, since the electron-phonon scattering
has also a pair-breaking component at finite temperatures
$T$. However, the mechanism of Mikhailovsky et al. predicts that
$\Gamma$ scales with $T$ according to $\Gamma\propto T^3$, and
therefore it is not of direct relevance to the experiments of
\cite{Szabo16,White86,Sherman15} and the like, where the parameter
$\Gamma$ is only weakly $T$-dependent and does not vanish in the
low-temperature limit.

The explanation of Eq.~(\ref{eq:dynes_delta}) should be therefore
sought in presence of elastic pair-breaking processes, such as
scattering on magnetic impurities \cite{Abrikosov61} and/or
fluctuating order parameter \cite{Feigelman12}. However, the latter
possibility seems to be ruled out by the spatial homogeneity of the
tunneling spectra observed in \cite{Szabo16}. Moreover, a fluctuating
order parameter is expected to produce appreciable change of
$N(\omega)$ only for $|\omega|\approx {\bar \Delta}$
\cite{Feigelman12}. Therefore we will concentrate only on the effect
of magnetic impurities.

\begin{figure}[t]
\includegraphics[width=6.5cm]{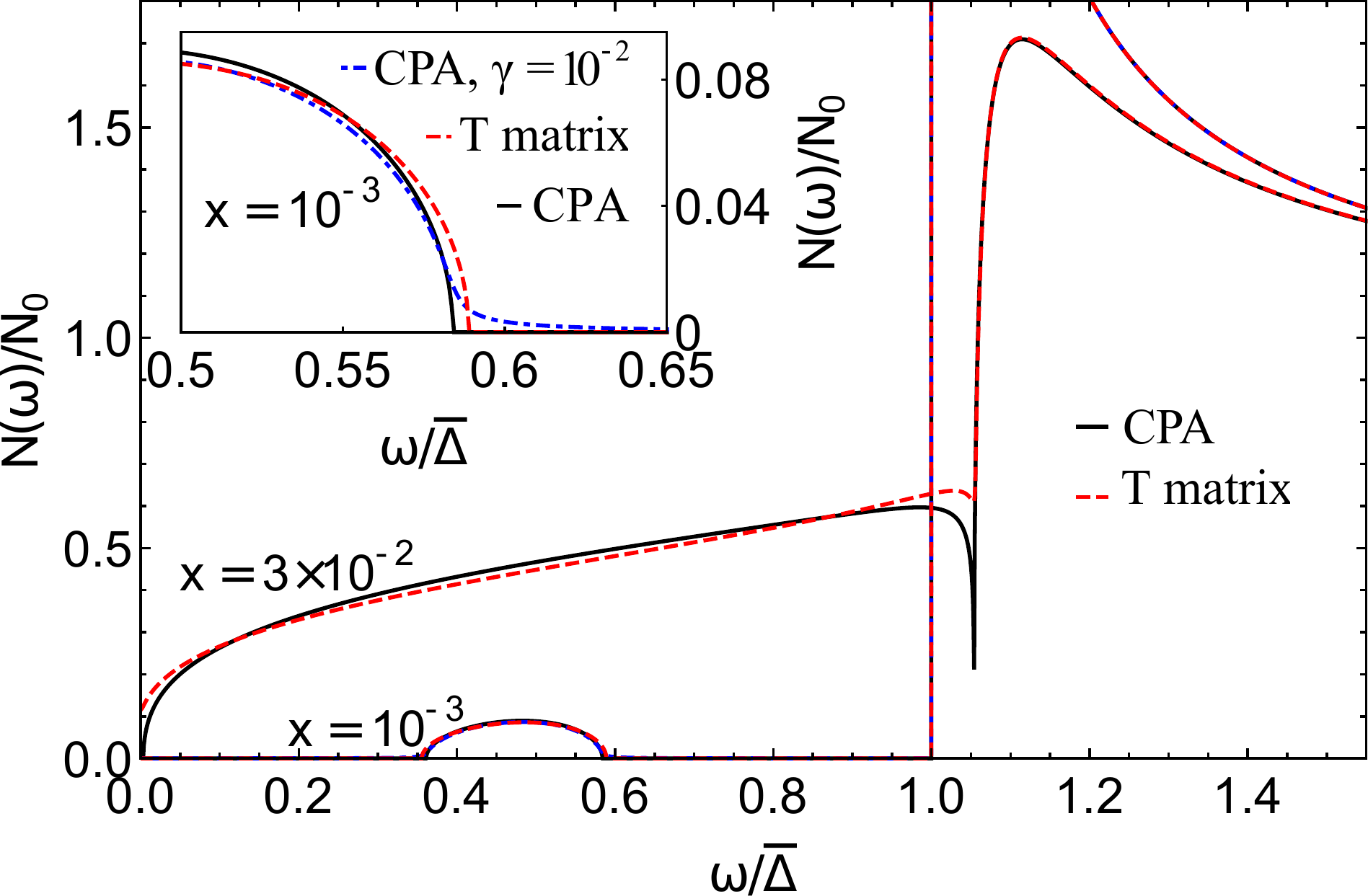}
\caption{(Color online) $N(\omega)$ for a superconductor with dilute
  pair-breaking impurities with $\lambda_0=0.6$ and $\pi N_0{\bar
    \Delta}=0.05$. Results for two impurity concentrations are shown,
  $x=0.001$ and $x=0.03$. The inset shows that, within CPA, the hard
  spectral edge of the impurity band softens if we replace the delta
  functions in Eq.~(\ref{eq:impurity}) by Lorentzians with widths
  $\gamma V_0$.}
\label{fig:tmatrix}
\end{figure}

It should be pointed out that, when the magnetic impurities are
treated in the Born approximation \cite{Abrikosov61}, the functional
form Eq.~(\ref{eq:dynes_delta}) does arise, but only in the limit
$\Gamma\gg {\bar \Delta}$, which is not of direct relevance to the
data in Fig.~\ref{fig:szabo}.  Subsequent theoretical work which went
beyond the Born approximation concentrated on the limit of dilute
magnetic impurities.  Within the T-matrix approximation, which should
be essentially exact in the dilute impurity limit, Shiba has found
magnetic impurity-induced bound states inside the energy gap in
absence of additional potential disorder \cite{Shiba68}, and the
precise energy of such bound states was found to depend on the
coupling strength to the impurities. Furthermore, finite concentration
of magnetic impurities was shown to lead to the formation of impurity
bands centered at the bound-state energies, see
Fig.~\ref{fig:tmatrix}. Provided the magnetic impurities are dilute,
later it was shown that presence of additional strong potential
disorder does not change these results \cite{Marchetti02}, and very
recently it has been argued that even going beyond mean-field theory
leads to only marginal changes of Shiba's results \cite{Fominov15}.

It seems to be clear then that, in order to reproduce
Eq.~(\ref{eq:dynes}) in the physically relevant case $\Gamma\lesssim
{\bar \Delta}$, one has to allow for spatially varying coupling
strengths to impurities, but in such a way which leads to a spatially
uniform gap function. This forces us to allow for a dense distribution
of impurities and therefore we have to abandon the previously used
techniques \cite{Shiba68,Marchetti02,Fominov15}. In this paper we have
chosen to make use of the coherent potential approximation (CPA),
which is well known to provide a successful description of
single-particle properties in disordered systems
\cite{Soven67,Velicky68,Weinkauf75}.

{\it CPA equations.} Within CPA we look for an averaged Nambu-Gorkov
Green's function $\hat{G}_M$ defined by
$\hat{G}_M^{-1}=\hat{G}_0^{-1}-\hat{\Sigma}$ where
$\hat{G}_0^{-1}({\bf k},\omega_n)=i\omega_n\tau_0-\varepsilon_{\bf
  k}\tau_3$ is the bare Green's function and
$\hat{\Sigma}_n=-i\Gamma_n\tau_0+\Phi_n\tau_1+\chi_n\tau_3$ is a local
translationally invariant self-energy generated by disorder and
pairing interactions. We work in imaginary time formalism, the index
$n$ denotes the Matsubara frequency, and $\tau_i$ are the Pauli
matrices.  

For the impurity potential we take 
$$
\hat{V}={\bar \Delta}\tau_1+U\tau_3+V\tau_0.
$$  
The first term is the spatially homogeneous pairing interaction, the
second term is a fluctuating potential which is usually large in
samples described by Eq.~(\ref{eq:dynes}), and the last term is a much
weaker classical pair-breaking field, polarized along a fixed
direction in spin space \cite{Bzdusek15}. We assume that the fields
$U$ and $V$ are distributed according to independent and spatially
uncorrelated even functions $P_s(U)$ and $P_m(V)$.

In CPA the self-energy $\hat{\Sigma}$ is chosen so that, on average,
electrons described by $\hat{G}_M$ do not scatter on the random
potential $\hat{V}$. This leads to the self-consistent equation
for the self-energy \cite{SM},
\begin{equation}
\left\langle(\hat{V}-\hat{\Sigma})
\left[{\bf 1}-\hat{G}_{\rm loc}(\hat{V}-\hat{\Sigma})\right]^{-1}
\right\rangle_{U,V}=0,
\label{eq:cpa_def}
\end{equation}
where the angular brackets denote averaging with respect to $U$, $V$
and $\hat{G}_{\rm loc}=(\hat{G}_M)_{ii}$ is the diagonal component (in
coordinate space) of $\hat{G}_M$.

For a particle-hole symmetric system, the defining
Eq.~(\ref{eq:cpa_def}) of CPA is compatible with $\chi_n=0$ \cite{SM}.
In what follows we use dimensionless pair-conserving and pair-breaking
fields $\mu=\pi N_0 U$ and $\lambda=\pi N_0 V$, respectively.  For
convenience, we also make use of the dimensionless quantities
$\gamma_n=\pi N_0\Gamma_n$, $\Lambda_n=\lambda+i\gamma_n$, and
$\delta_n=\pi N_0({\bar \Delta}-\Phi_n)$, as well as of the auxiliary
variables
\begin{eqnarray*}
z_n=x_n+i y_n=
\left[\Phi_n+i(\omega_n+\Gamma_n)\right]/
\sqrt{(\omega_n+\Gamma_n)^2+\Phi_n^2}
\end{eqnarray*}
which satisfy the identity $|z_n|^2=1$. In terms of these variables,
Eq.~(\ref{eq:cpa_def}) can be rewritten as a single complex equation
\cite{SM},
\begin{eqnarray}
\left\langle\frac{z_n+\delta_n-\Lambda_n}
{(z_n+\delta_n-\Lambda_n)(z_n^\ast+\delta_n+\Lambda_n)+\mu^2}
\right\rangle_{\mu,\lambda}=z_n.
\label{eq:cpa}
\end{eqnarray}
By solving Eq.~(\ref{eq:cpa}), we can find the normal and anomalous
self-energies $\Gamma_n$ and $\Phi_n$, or, alternatively, the
wave-function renormalization $Z_n=1+\Gamma_n/\omega_n$ and the gap
function $\Delta_n=\Phi_n/Z_n$.

{\it Dilute gas of identical magnetic impurities.} In order to
proceed, we need to specify the probability distributions $P_s(U)$ and
$P_m(V)$. We will start by considering the well studied example with
vanishing potential disorder and
\begin{equation}
P_m(V)=(1-x)\delta(V)
+\frac{x}{2}\left[\delta(V-V_0)+\delta(V+V_0)\right],
\label{eq:impurity}
\end{equation}
which describes a set of magnetic impurities with magnetic field $\pm
V_0$ and concentration $x$. Making use of this distribution in
Eq.~(\ref{eq:cpa}) and assuming that $x\ll 1$, to first order in the
impurity concentration we find
\begin{eqnarray*}
Z_n&=&1+
\frac{\Gamma_0(1+\lambda_0^2)\sqrt{\omega_n^2+\Delta_n^2}}
{(1+\lambda_0^2)^2\omega_n^2+(1-\lambda_0^2)^2\Delta_n^2},
\\
{\bar \Delta}&=&
\left[1+2\Gamma_0\frac{\sqrt{\omega_n^2+\Delta_n^2}}
{(1+\lambda_0^2)^2\omega_n^2+(1-\lambda_0^2)^2\Delta_n^2}\right]\Delta_n,
\end{eqnarray*}
where $\Gamma_0=x \pi N_0V_0^2$ and $\lambda_0=\pi
N_0V_0$. These are the well-known self-consistent equations of the
T-matrix approximation \cite{Shiba68}, which shows that CPA becomes
exact in the low-density limit.

In Fig.~\ref{fig:tmatrix} we compare $N(\omega)$ for a superconductor
with a dilute gas of pair-breaking impurities, calculated within the
T-matrix approximation and the full CPA. Both approximations result in
a qualitatively similar density of states.  As expected, the agreement
between the two approximations improves as the impurity concentration
$x$ decreases. Somewhat surprisingly, CPA predicts systematically
narrower impurity bands.

{\it The Dynes superconductors.} Now we turn to the main result of
this paper. In order to take into account the spatial distribution of
coupling strengths to magnetic impurities, instead of
Eq.~(\ref{eq:impurity}) we consider the so-called Lloyd model
\cite{Lloyd69}, $P_m(V)=\pi^{-1}\Gamma/(V^2+\Gamma^2)$, with a
continuous spread of impurity strengths ranging up to $\sim
\Gamma$. We emphasize that we don't need to make any further
assumptions about $P_s(U)$.

Let us for definiteness consider $\omega_n>0$ and assume that
$y_n>\gamma_n>0$.  Inserting $P_m(V)$ into Eq.~(\ref{eq:cpa}), we
notice that averaging with respect to $\lambda$ can be readily
performed in the complex plane of $\lambda$, leading to
\begin{equation}
\left\langle \zeta_n/(|\zeta_n|^2+\mu^2)\right\rangle_\mu=z_n,
\label{eq:anderson}
\end{equation}
where we have introduced
$\zeta_n=(x_n+\delta_n)+i(y_n+\lambda_0-\gamma_n)$ with $\lambda_0=\pi
N_0\Gamma$. Comparing the phases of both sides of
Eq.~(\ref{eq:anderson}) leads to $\Delta_n=\omega_n{\bar
  \Delta}/(\omega_n+\Gamma)$ or, after analytic continuation to the
real axis, to Eq.~(\ref{eq:dynes_delta}).  This means that, within
CPA, the Lorentzian distribution $P_m(V)$ of pair-breaking fields
generates precisely that frequency-dependent gap function
$\Delta(\omega)$ which reproduces the Dynes tunneling density of
states Eq.~(\ref{eq:dynes}).  Moreover, the Dynes parameter $\Gamma$
is given directly by the width of the Lorentzian $P_m(V)$.  Note that
in absence of pair breaking, i.e. for $\Gamma=0$, CPA predicts
$\Delta(\omega)={\bar \Delta}$, which is consistent with the Anderson
theorem.

Comparing the amplitudes of both sides of Eq.~(\ref{eq:anderson}) we
find that $|\zeta_n|=F$ is independent of frequency and the constant
$F$ is fixed by $\int d\mu P_s(\mu)F/(\mu^2+F^2)=1$.  The self-energy
$\Gamma_n$ can be determined from $|\zeta_n|=F$.  After analytic
continuation to the real axis the wave-function renormalization
$Z(\omega)=1+i\Gamma(\omega)/\omega$ reads
\begin{equation}
Z(\omega)=(1+i\Gamma_s/\Omega)(1+i\Gamma/\omega),
\label{eq:dynes_gamma}
\end{equation}
where $\Gamma_s=(1-F)/\pi N_0$ is the pair-conserving scattering rate
and $\Omega=\left[(\omega+i\Gamma)^2-{\bar \Delta}^2\right]^{1/2}$.
The function $Z(\omega)$ is seen to be a product of two factors. The
first factor, due to pair-conserving scattering, reproduces the Born
approximation \cite{Zhu04}, albeit with a generalized $\Gamma_s$.  The
second factor, due to pair-breaking processes, has the same form as
found previously for inelastic processes at finite temperatures
\cite{Mikhailovsky91}. Strongly disordered samples which we are
interested in are described by $\Gamma\lesssim{\bar \Delta}\ll
\Gamma_s$.

The criterion for applicability of our results, $y_n>\gamma_n$, is
satisfied for $F>g=\pi N_0 (\Gamma^2+{\bar \Delta}^2)^{1/2}$.  If for
$P_s(U)$ we take, as an order-of-magnitude estimate, a box
distribution of width $2U_0$, we find $F=\pi N_0 U_0/\tan(\pi N_0
U_0)$. On the other hand, for samples with $\Gamma\lesssim{\bar
  \Delta}$ we have $g\ll 1$.  From here it follows that $F>g$ holds
provided that $U_0\lesssim 1/(2N_0)$, i.e. up to large potential
disorder \cite{SM}.

We emphasize that our microscopics goes beyond the phenomenology of
Eq.~(\ref{eq:dynes}) by predicting {\it both} of the Eliashberg
functions, $\Delta(\omega)$ and $Z(\omega)$.  The resulting retarded
electron Green's function reads
\begin{equation}
\hat{G}_M({\bf k},\omega)=
\frac{(1+i\Gamma_s/\Omega)\left[(\omega+i\Gamma)\tau_0
+{\bar \Delta}\tau_1\right]
+\varepsilon_{\bf k}\tau_3}
{(\Omega+i\Gamma_s)^2-\varepsilon_{\bf k}^2}.
\label{eq:dynes_green}
\end{equation}
Note that Eq.~(\ref{eq:dynes_green}) is the simplest consistent
generalization of the BCS Green's function which takes into account
both, the pair-conserving and the pair-breaking scattering processes
with rates $\Gamma_s$ and $\Gamma$, respectively. Superconductors
described by Eq.~(\ref{eq:dynes_green}) will be called Dynes
superconductors in what follows.

\begin{figure}[t]
\includegraphics[width=6.5cm]{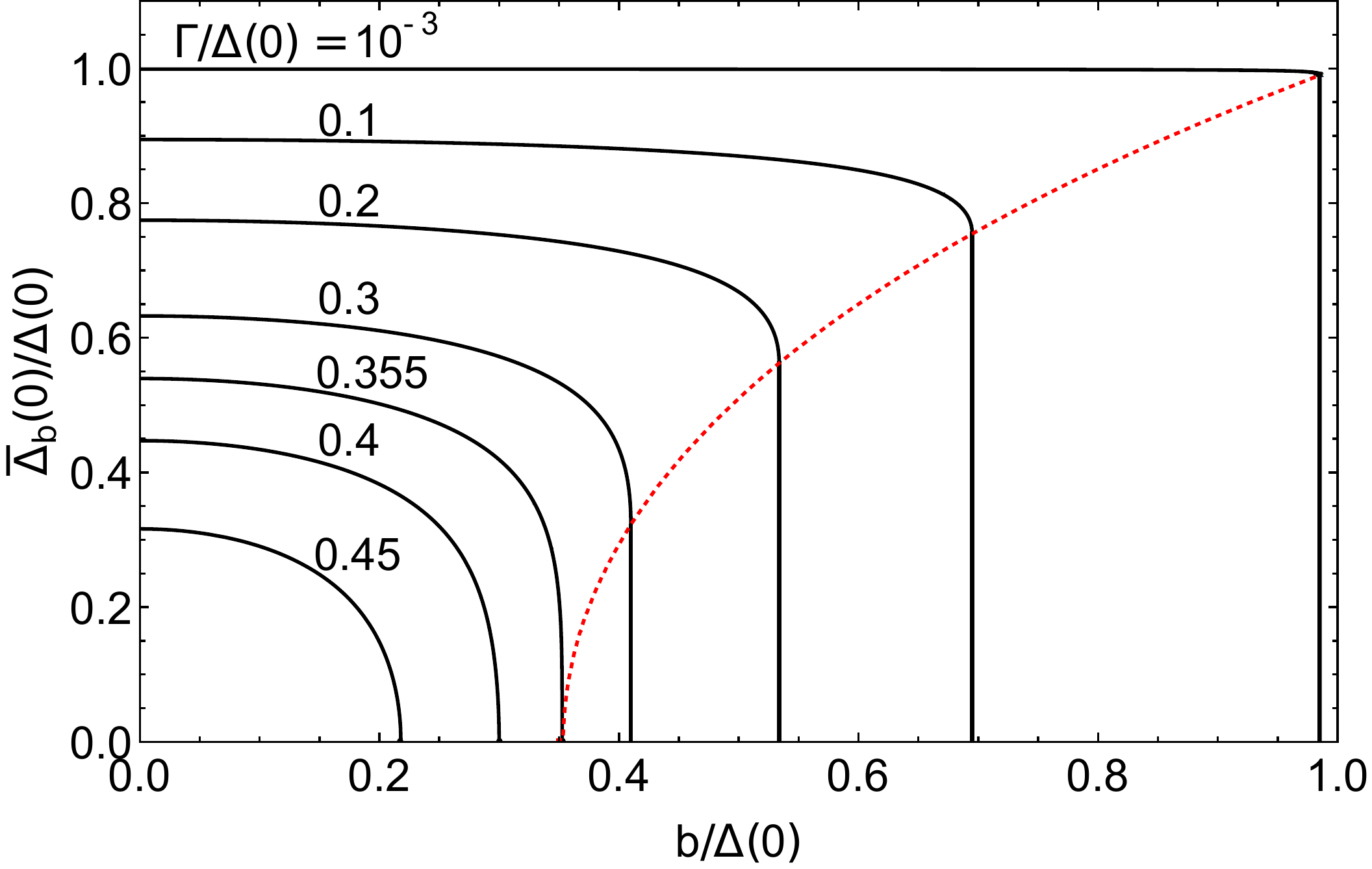}
\caption{(Color online) The order parameter at $T=0$ in a magnetic
  field $b$, $\bar{\Delta}_b(0)$, as a function of $b$ for several
  $\Gamma$. First-order transitions for small $\Gamma$ are shown by
  the dotted line.}
\label{fig:thermo_B}
\end{figure}

{\it Thermodynamics.} Next we consider the thermodynamic properties of
the Dynes superconductors. To this end, we realize that the
off-diagonal part ${\bar\Delta}$ of the potential ${\hat V}$ has to
come from a phonon-induced anomalous self-energy. Therefore, within
the BCS approximation with dimensionless coupling constant $\lambda\ll
1$ and cut-off frequency $\Omega$, we find the self-consistent
equation \cite{SM}
\begin{equation}
{\bar \Delta}=\lambda\pi T\sum_{\omega_n=-\Omega}^\Omega
\left[
{\bar \Delta}/\sqrt{(|\omega_n|+\Gamma)^2+{\bar \Delta}^2}
\right].
\label{eq:selfconsistent}
\end{equation}
Making use of Eq.~(\ref{eq:selfconsistent}), we can calculate the
temperature dependence $\bar{\Delta}=\bar{\Delta}(T)$ as a function of
the parameter $\Gamma$.  We find that the critical temperature of a
dirty Dynes superconductor, ${\bar T}_c$, is governed by the same
equation as in the Abrikosov-Gorkov theory,
$\psi(\tfrac{1}{2}+\tfrac{\alpha}{x})-\psi(\tfrac{1}{2})
=\ln(\tfrac{1}{x})$, where $\psi(x)$ is the digamma function,
$\alpha=\Gamma/(2\pi T_c)$, $x={\bar T}_c/T_c$, and $T_c$ is the
critical temperature of the clean system. This is because, as already
mentioned, close to the critical temperature,
Eq.~(\ref{eq:dynes_delta}) applies to superconductors with pair
breaking even in the Born approximation.

Below $T_c$ it is convenient to normalize $\bar{\Delta}(T)$ in terms
of $\Delta(0)$, the zero-temperature gap of the clean system. At $T=0$
we find $\bar{\Delta}(0)=\sqrt{\Delta(0)[\Delta(0)-2\Gamma]}$,
therefore the critical disorder strength for complete disappearance of
superconductivity is $\Gamma_c=\Delta(0)/2$. The $\bar{\Delta}=
\bar{\Delta}(T)$ curves for varying $\Gamma$ are essentially BCS-like
\cite{SM}.  The ratio ${\bar \Delta}(0)/{\bar T_c}$ increases by a
factor ${\cal R}$ with respect to the clean-system value
$\Delta(0)/T_c$ and ${\cal R}$ slightly grows with $\Gamma$. For
$\Gamma\rightarrow \Gamma_c$ we find ${\cal R}(\Gamma_c)\approx 1.45$,
which is however much less than ${\cal R}(\Gamma_c)\approx 2.52$
within the Abrikosov-Gorkov theory.

\begin{figure}[t]
\includegraphics[width=8.5cm]{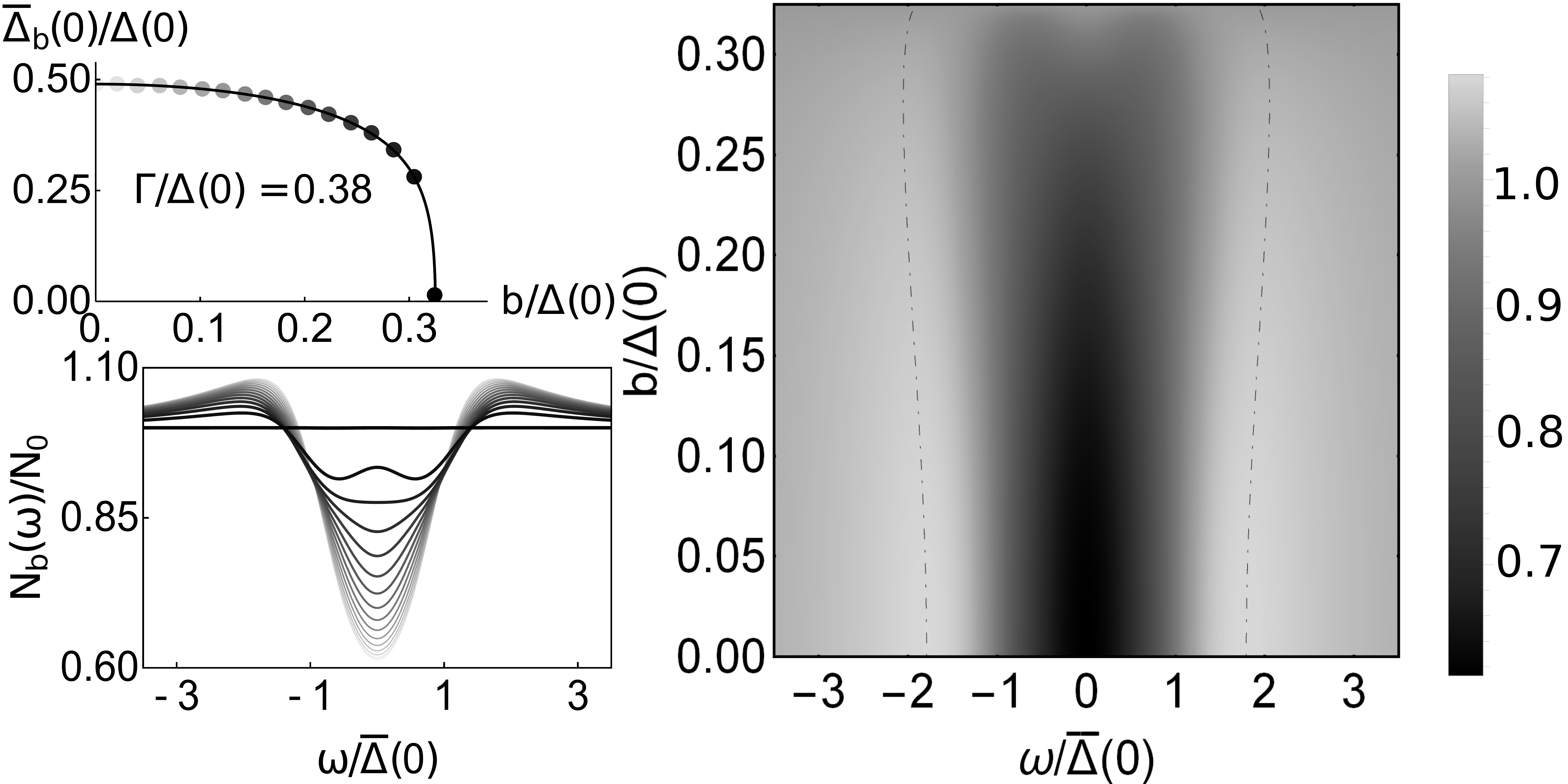}
\caption{Right panel: map of $N_b(\omega)$ at $T=0$ in the
  $(\omega,b)$ plane for a Dynes superconductor with
  $\Gamma/\Delta(0)=0.38$. The dash-dotted curve marks the positions
  of the maxima of $N_b(\omega)$ at fixed $b$. The lower left panel
  shows $N_b(\omega)$ for several values of $b$.  The self-consistent
  values of ${\bar \Delta}_b(0)$ for the same $b$-values are plotted
  in the upper left panel.}
\label{fig:dos_B}
\end{figure}

{\it Effect of external magnetic field.} Finally we study the density
of states of a Dynes superconductor in an external magnetic field
$B$. We assume that the superconductor is sufficiently dirty, so that
the suppression of $\bar{\Delta}$ by $B$ can be roughly estimated by
keeping only the Zeeman coupling \cite{SM}.  In this approximation the
effect of $B$ is fully described by simply changing the bare electron
Green's function to $\hat{G}_0^{-1}({\bf
  k},\omega_n)=(i\omega_n-b)\tau_0-\varepsilon_{\bf k}\tau_3$ with
$b=\mu_BB$. One can check that the CPA expressions remain valid, if we
make the substitution $\omega_n\rightarrow \omega_n+ib$. In
particular, Eq.~(\ref{eq:selfconsistent}) is replaced by the following
self-consistent equation,
\begin{equation}
{\bar \Delta}=2\lambda\pi T\sum_{\omega_n>0}^\Omega
{\rm Re}\left[
{\bar \Delta}/\sqrt{(\omega_n+\Gamma+ib)^2+{\bar \Delta}^2}
\right].
\label{eq:magfield}
\end{equation}
As was to be expected, the theory with only Zeeman coupling,
Eq.~(\ref{eq:magfield}), predicts a first-order transition at small
$\Gamma$, but, as shown in Fig.~\ref{fig:thermo_B}, the transition
becomes continuous for $\Gamma>\Gamma_c\approx 0.355\Delta(0)$, as one
would expect in the full theory with orbital effects included.

Sufficiently far away from the vortex cores, the density of states in
a finite magnetic field $N_b(\omega)$ can be described by considering
only the Zeeman coupling \cite{SM}, and this leads to
$N_b(\omega)=\sum_{\pm}N(\omega\pm b)/2$. In Fig.~\ref{fig:dos_B} we
plot the evolution of $N_b(\omega)$ with $b$ for a Dynes
superconductor with $\Gamma/\Delta(0)=0.38$. Due to the Zeeman
coupling, the peak-to-peak distance of the density of states exhibits
only small changes with $b$, up to the critical field $b_c$. This
means that gap filling rather than gap closing with increasing $b$ can
be observed in dirty Dynes superconductors. Note, however, that the
order parameter $\bar{\Delta}_b(0)$ does behave in a standard way and
vanishes at $b_c$, see the left panel of Fig.~\ref{fig:dos_B}.  Very
recently, similar behavior of $N(\omega)$ in magnetic fields has been
observed experimentally \cite{Szabo17}.

{\it Conclusions.} We have identified a class of gapless
superconductors, the Dynes superconductors, which are distinguished by
a sufficiently broad distribution of pair-breaking fields.  The Dynes
superconductors are described by two scattering rates, $\Gamma_s$ and
$\Gamma$, for pair-conserving and pair-breaking processes,
respectively.  The Green function of a canonical Dynes superconductor
is given by Eq.~(\ref{eq:dynes_green}). We have shown that this
functional form follows from the CPA equations with a Lorentzian
distribution of pair-breaking fields and arbitrary potential
disorder. The Dynes superconductors are always gapless from $T_c$ all
the way down to the lowest temperatures, and their thermodynamic
properties differ from predictions of the Abrikosov-Gorkov theory.

We thank P. Szab\'{o} for sending us the data presented in
Fig.~\ref{fig:szabo}, and P.~Marko\v{s}, P.~Samuely and M.~Grajcar for
discussions. This work was supported by the Slovak Research and
Development Agency under contract No.~APVV-0605-14 and by the Agency
VEGA under contract No.~1/0904/15.

\end{document}


\title{SUPPLEMENTARY MATERIAL\\
Microscopic interpretation of the Dynes formula for the tunneling
  density of states}

\author{Franti\v{s}ek Herman and Richard Hlubina}
\affiliation{Department of Experimental Physics, Comenius
  University, Mlynsk\'{a} Dolina F2, 842 48 Bratislava,
  Slovakia}
\maketitle

\section{Coherent potential approximation}
For convenience we present a short sketch of the derivation of Eq.(4)
from the main text. Let $\hat{G}$ be the full Green's function of the
disordered system (i.e. a matrix whose indices describe the lattice
sites and the Nambu components) and let $\hat{G}_0$ be the bare
Green's function of the clean system. Then $\hat{G}$ satisfies the
matrix equation
\begin{equation}
\hat{G}=\hat{G}_0+\hat{G}_0\hat{V}\hat{G},
\label{eq:green_0}
\end{equation}
which describes repeated scattering of electrons described by
$\hat{G}_0$ by the the random potential $\hat{V}$. Equivalently,
Eq.~(\ref{eq:green_0}) can be written in terms of the T-matrix
$\hat{T}_0$ in the form
$\hat{G}=\hat{G}_0+\hat{G}_0\hat{T}_0\hat{G}_0$.  Comparing these two
expressions for $\hat{G}$, one finds easily that
\begin{equation}
\hat{T}_0=\hat{V}({\bf 1}-\hat{G}_0\hat{V})^{-1}.
\label{eq:tmatrix_0}
\end{equation}

In CPA we look for an optimal averaged Nambu-Gorkov Green's function
$\hat{G}_M$ describing the disordered medium.  Let us express this
effective Green's function in terms of the self-energy by the Dyson
equation
\begin{equation}
\hat{G}_M^{-1}=\hat{G}_0^{-1}-\hat{\Sigma}.
\label{eq:dyson}
\end{equation}
From similar considerations which led to Eq.~(\ref{eq:green_0}) it
follows that the full Green's function of the disordered system
$\hat{G}$ satisfies also the matrix equation
\begin{equation}
\hat{G}=\hat{G}_M+\hat{G}_M(\hat{V}-\hat{\Sigma})\hat{G},
\label{eq:green_M}
\end{equation}
which shows that electrons described by the effective Green's function
$\hat{G}_M$ interact with a reduced potential
$\hat{V}-\hat{\Sigma}$. 

In order to fix the optimal self-energy $\hat{\Sigma}$, let us rewrite
Eq.~(\ref{eq:green_M}) for the full Green's function $\hat{G}$ in
terms of the T-matrix of the effective medium $\hat{T}$ by
$\hat{G}=\hat{G}_M+\hat{G}_M\hat{T}\hat{G}_M$. A calculation
completely analogous to that leading to Eq.~(\ref{eq:tmatrix_0}) leads
then to an expression for the T-matrix of the effective medium:
\begin{equation}
\hat{T}=(\hat{V}-\hat{\Sigma})
\left[{\bf 1}-\hat{G}_M(\hat{V}-\hat{\Sigma})\right]^{-1}.
\label{eq:tmatrix_M}
\end{equation}
Note that Eq.~(\ref{eq:tmatrix_M}) differs from
Eq.~(\ref{eq:tmatrix_0}) by simply replacing $\hat{V}$ by
$\hat{V}-\hat{\Sigma}$ and $\hat{G}_0$ by $\hat{G}_M$, i.e. $\hat{T}$
describes residual scattering on disorder, not taken into account in
the effective medium description. Two points are to be noted: (i)
$\hat{T}$ for a given sample depends on the choice of the random
potential, and (ii) $\hat{T}$ is a matrix in the coordinate space.

Now it is natural to choose the effective medium so that, after
averaging over disorder, the residual scattering is minimized,
$\langle\hat{T}\rangle=0$. Within CPA one requires that only the
site-diagonal components of the T-matrix vanish. This leads to the
self-consistent equation \cite{Velicky68}
\begin{equation}
\left\langle(\hat{V}-\hat{\Sigma}) \left[{\bf 1}-\hat{G}_{\rm
    loc}(\hat{V}-\hat{\Sigma})\right]^{-1} \right\rangle=0,
\label{eq:cpa_def}
\end{equation}
where $\hat{G}_{\rm loc}=(\hat{G}_M)_{ii}$ is the diagonal component
(in coordinate space) of $\hat{G}_M$. This is Eq.(4) from the main
text.

\section{Derivation of Eq.(5)}
Let us take for the self-energy the ansatz
$\hat{\Sigma}_n=-i\Gamma_n\tau_0+\Phi_n\tau_1+\chi_n\tau_3$ from the
main text and making use of Eq.~(\ref{eq:dyson}) let us calculate the
averaged Green's function $\hat{G}_{M}({\bf k},\omega_n)$.  We find
\begin{equation}
\hat{G}_{M}({\bf k},\omega_n)=-\frac
{i(\omega_n+\Gamma_n)\tau_0+\Phi_n\tau_1+(\varepsilon_{\bf k}+\chi_n)\tau_3}
{(\omega_n+\Gamma_n)^2+\Phi_n^2+(\varepsilon_{\bf k}+\chi_n)^2}.
\label{eq:green}
\end{equation}
The local Green's function $\hat{G}_{\rm loc}(\omega_n)$ can be found
by Fourier transforming the function $\hat{G}_{M}({\bf k},\omega_n)$
from momentum~(${\bf k}$) to real~(${\bf r}$) space and by taking
${\bf r}=0$.  Replacing the momentum summation by energy integration
and assuming a constant density of states $N_0$ in the vicinity of the
Fermi level, a standard calculation leads to
\begin{equation}
\hat{G}_{\rm loc}(\omega_n) = -\pi N_{0}
\frac{i(\omega_n + \Gamma_n)\tau_0 + \phi_n\tau_1}
{\sqrt{(\omega_n + \Gamma_n)^2 + \phi_n^2}}.
\label{eq:green_local}
\end{equation}
Note that, as usual, the component of $\hat{G}_{\rm loc}$ proportional
to the Pauli matrix $\tau_3$ vanishes. This is a consequence of the
assumed particle-hole symmetry of the problem.

Evaluating the matrix inverse entering Eq.~(\ref{eq:cpa_def}) is
straightforward, since $\hat{G}_{\rm loc}$, $\hat{V}$, and
$\hat{\Sigma}$ are matrices $2\times2$.  Making use of the explicit
form of the potential $\hat{V}={\bar \Delta}\tau_1+U\tau_3+V\tau_0$
from the main text we find
\begin{equation}
\left[\tau_0 - \hat{G}_{\rm loc}(\hat{V} - \hat{\Sigma})\right]^{-1} = 
\frac{a_n\tau_0 + i b_n \tau_1 + i c_n \tau_2 + i d_n \tau_3}
{a_n^2 + b_n^2 + c_n^2 + d_n^2},
\label{eq:inversion}
\end{equation}
where we have introduced auxiliary variables
\begin{align*}
a_n &= 1 + \pi N_0 
\frac{i\left(\omega_n+\Gamma_n\right)(V + i\Gamma_n) 
+ \phi_n(\bar{\Delta}-\phi_n)}
{\sqrt{(\omega_n + \Gamma_n)^2 + \phi_n^2}},
\\
b_n &= \pi N_0 
\frac{i\phi_n(V + i\Gamma_n) 
- \left(\omega_n + \Gamma_n\right)(\bar{\Delta} - \phi_n)}
{\sqrt{(\omega_n + \Gamma_n)^2 + \phi_n^2}},
\\
c_n &= \pi N_0 \frac{\phi_n(U - \chi_n)}
{\sqrt{(\omega_n + \Gamma_n)^2 + \phi_n^2}},\\
d_n &= \pi N_0 \frac{(\omega_n + \Gamma_n)(\chi_n - U)}
{\sqrt{(\omega_n + \Gamma_n)^2 + \phi_n^2}}.
\end{align*}
Inserting the result Eq.~(\ref{eq:inversion}) into
Eq.~(\ref{eq:cpa_def}), we obtain 4 equations, which follow from
requiring that the coefficients in front of the Pauli matrices
$\tau_i$ with $i=0,\ldots,3$ vanish:
\begin{align}
\left\langle
\frac{(V + i\Gamma_n)a_n + i(\bar{\Delta} - \phi_n)b_n + i(U - \chi_n)d_n}
{a_n^2+b_n^2+c_n^2+d_n^2} \right\rangle= 0, 
\label{tau0}\\
\left\langle
\frac{i(V + i\Gamma_n)b_n + (\bar{\Delta} - \phi_n)a_n + (U - \chi_n)c_n}
{a_n^2+b_n^2+c_n^2+d_n^2} \right\rangle= 0, 
\label{tau1}\\
\left\langle
\frac{i(V + i\Gamma_n)c_n + (\bar{\Delta} - \phi_n)d_n - (U - \chi_n)b_n}
{a_n^2+b_n^2+c_n^2+d_n^2} \right\rangle= 0, 
\label{tau2}\\
\left\langle
\frac{i(V + i\Gamma_n)d_n - (\bar{\Delta} - \phi_n)c_n + (U - \chi_n)a_n}
{a_n^2+b_n^2+c_n^2+d_n^2} \right\rangle= 0. 
\label{tau3}
\end{align}

If one makes use of the explicit form of the auxiliary variables
$a_n$, $b_n$, $c_n$ and $d_n$, the last two
equations~(\ref{tau2},\ref{tau3}) can be easily solved. In fact,
Eq.~\eqref{tau2} is trivially satisfied, and Eq.~\eqref{tau3} can be
written as
\begin{equation*}
\left\langle \frac{U - \chi_n}
{a_n^2 + b_n^2 + \left[\pi N_0(U-\chi_n)\right]^2} \right\rangle = 0.
\end{equation*}
Note that the variables $a_n$ and $b_n$ do not include the scalar
potential $U$. But since the distribution function $P(U)$ is supposed
to be even, one checks easily that Eq.~\eqref{tau3} is solved by
requiring $\chi_n = 0$.

Finally, if we take the sum and the difference of
Eqs.~(\ref{tau0},\ref{tau1}) and if we make use of the result $\chi_n
= 0$, we obtain another set of two equations. They can be written down
in a simple form by using the dimensionless variables $\mu$,
$\Lambda_n$, $\delta_n$ and $z_n$ defined in the main text:
\begin{align*}
\left\langle
\frac{(\delta_n + \Lambda_n)
\big(1 + z_n^{*}(\delta_n - \Lambda_n)\big) + z_n^*\mu^2}
{\big(1 + z_n^{*}(\delta_n - \Lambda_n)\big)
\big(1 + z_n(\delta_n + \Lambda_n)\big) + \mu^2}\right\rangle &= 0,
\\
\left\langle
\frac{(\delta_n - \Lambda_n)
\big(1 + z_n(\delta_n + \Lambda_n)\big) + z_n\mu^2}
{\big(1 + z_n^{*}(\delta_n - \Lambda_n)\big)
\big(1 + z_n(\delta_n + \Lambda_n)\big) + \mu^2}\right\rangle &= 0. 
\end{align*}
Assuming that $\phi_n$ and $\Gamma_n$ are purely real, we can easily
see that they reduce to just one equation after complex conjugation
and substitution $V\rightarrow -V$ in one of them. After some trivial
algebra we are therefore left with just one complex integral CPA
equation in the form of Eq.(5) from the main text.

\section{CPA in the normal state}
In the normal state our model for disorder implies that electrons with
spin $\sigma$ experience a random potential $W=U+\sigma V$ with
distribution functions
\begin{equation}
P_\sigma(W)=\int dU \int dV P_s(U) P_m(V) \delta(U+\sigma V-W).
\label{eq:distribution}
\end{equation}
Note that since $P_m(V)$ is even, we have
$P_\uparrow(W)=P_\downarrow(W)\equiv P(W)$. In the upper half-plane
$\omega_n>0$, Eq.~(4) from the main text is solved for this
distribution function by a frequency-independent self-energy
$\Sigma_n=-i\Gamma_N$, where $\Gamma_N=(1-F_N)/(\pi N_0)$ and the
constant $F_N$ is given by
\begin{equation}
1=\left\langle
\frac{F_N}{F_N^2+(\pi N_0W)^2}
\right\rangle_W .
\label{eq:normal_F}
\end{equation}
Note that Eq.~(\ref{eq:normal_F}) does not have a solution for
sufficiently broad distributions $P(W)$. This is an artifact of the
CPA, as can be shown readily, if we take for $P_s(U)$ and $P_m(V)$
Lorentzians with widths $\Gamma_s$ and $\Gamma$, respectively. In
fact, in that case also $P(W)$ is a Lorentzian with width
$\Gamma+\Gamma_s$ and Eq.~(\ref{eq:normal_F}) implies that $1-F_N=\pi
N_0(\Gamma+\Gamma_s)$, or, in other words, the normal-state
self-energy is given by the width of $P(W)$,
$\Gamma_N=\Gamma+\Gamma_s$. However, since Eq.~(\ref{eq:normal_F})
clearly requires that $F_N>0$, the CPA solution is valid only for $\pi
N_0\Gamma_N<1$.

On the other hand, as shown by Lloyd \cite{Lloyd69}, the normal-state
model with a Lorentzian distribution $P(W)$ is exactly solvable for
all values of $\Gamma_N$, thus the criterion $\pi N_0\Gamma_N<1$ can
not have any physical meaning and it must be an artifact of the CPA.
It should be pointed out, however, that in its region of validity, the
CPA does reproduce the exact self-energy of the Lloyd model
\cite{Lloyd69}.

\section{Thermodynamics of the Dynes superconductors}
\begin{figure}[ht]
\includegraphics[width=7cm]{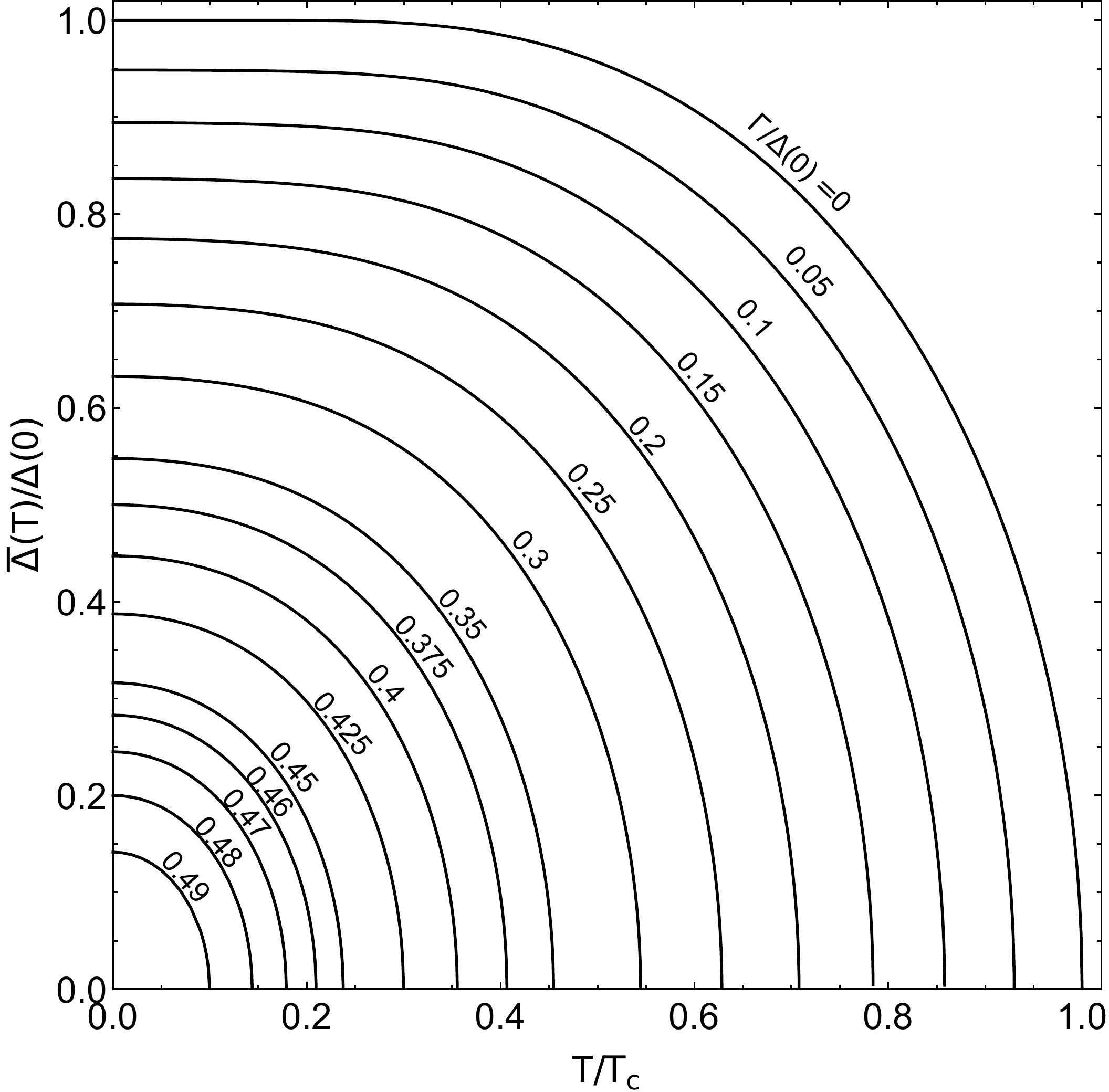}
\caption{Numerically determined ideal gaps ${\bar \Delta}(T)$ of Dynes
  superconductors for various pair-breaking parameters $\Gamma$ for
  fixed $\lambda\ll 1$ and $\Omega$. The gaps are measured in units of
  $\Delta(0)$, which is the gap of the clean system at
  $T=0$. Temperature is displayed in units of $T_c$, which is the
  critical temperature of the clean system. }
\label{fig:suppl_thermo}
\end{figure}

Let us assume that the pairing in the Dynes superconductors is driven
by a local phonon-mediated electron-electron interaction $U_{\rm ph}$
which is present up to a finite frequency cutoff $\Omega$. Then, at
the mean-field level, the off-diagonal part of the potential ${\hat
  V}$ is determined by the self-consistent equation
\begin{equation}
{\bar \Delta}=U_{\rm ph}
\langle\psi_\uparrow({\bf r})\psi_\downarrow({\bf r})\rangle,
\label{eq:delta_definition}
\end{equation}
where $\psi_\sigma({\bf r})$ are the annihilation operators for
electrons at site ${\bf r}$.  After Fourier transformation to momentum
space with annihilation operators $c_{{\bf k}\sigma}$, this equation
can be written as
\begin{equation}
{\bar \Delta}=\frac{U_{\rm ph}}{N}\sum_{\bf k}
\langle c_{{\bf k}\uparrow}c_{-{\bf k}\downarrow}\rangle
=-\frac{U_{\rm ph}}{N}\sum_{\bf k}\hat{G}_M^{12}({\bf k},\tau=0^+),
\label{eq:delta_fourier}
\end{equation} 
where $N$ is the number of lattice sites and $\hat{G}_M^{12}$ is the
off-diagonal component of the averaged Green's function.  Performing
the temporal Fourier transformation of the Green's function and making
use of the explicit form of $\hat{G}_M$, Eq.~(\ref{eq:green}),
together with the result $\chi_n=0$, Eq.~(\ref{eq:delta_fourier}) can
be written as
\begin{equation}
{\bar \Delta}= U_{\rm ph}\frac{T}{N}
\sum_{{\bf k},\omega_m} \frac{Z_m\Delta_m}
{Z_m^2(\omega_m^2+\Delta_m^2)+\varepsilon_{\bf k}^2}.
\label{eq:delta_fourier2}
\end{equation}
Let us note that the momentum summation in
Eq.~(\ref{eq:delta_fourier2}) can be replaced by energy integration,
which in turn can be performed explicitly. Imposing furthermore the
frequency cutoff $\Omega$, this leads to the result
\begin{equation}
{\bar \Delta}=\lambda\pi T\sum_{\omega_m=-\Omega}^\Omega
\frac{\Delta_m}{\sqrt{\omega_m^2+\Delta_m^2}},
\label{eq:self_consistent}
\end{equation}
where $\lambda=N_0U_{\rm ph}$ is a dimensionless coupling constant.
Note that the wave-function renormalization $Z_m$ drops out from the
right-hand side. If in Eq.~(\ref{eq:self_consistent}) we make use of
the frequency dependence of the gap function of a Dynes
superconductor, valid for both signs of $\omega_n$,
$$
\Delta_m=\frac{|\omega_m|}{|\omega_m|+\Gamma}{\bar \Delta},
$$ 
we finally end up with the self-consistent Eq.~(10) from the main
text.  It is worth pointing out that Eq.~(10) from the main text does
not contain the pair-conserving scattering rate $\Gamma_s$, and this
is consistent with the Anderson theorem. In
Fig.~\ref{fig:suppl_thermo} we show the temperature depence of the
ideal gaps ${\bar \Delta}(T)$ of Dynes superconductors for various
pair-breaking parameters $\Gamma$, which are seen to be essentially
BCS-like for all admissible values of $\Gamma$.

\section{Effect of finite external magnetic field}
External magnetic field interacts with electrons via two different
mechanisms: via the Zeeman coupling and by minimal coupling between
the electron's momentum and the vector potential, which for brevity
will be called orbital coupling.  In order to compare the relative
importance of the Zeeman and orbital couplings, we will estimate the
critical fields, i.e. those fields which lead to a complete
destruction of superconductivity, for both mechanisms taken
separately.  Let us start by considering the orbital coupling. In a
dirty type-II superconductor such as MoC, the upper critical field
$H_{c2}$ can be estimated as $\mu_0 H_{c2}\sim\Phi_0/(\xi_0 \ell)$,
where $\Phi_0$ is the flux quantum, $\xi_0\sim \hbar v_F/\Delta$ is
the coherence length, and $\ell$ is the mean free path. On the other
hand, due to the Zeeman coupling, the Cooper pairing will be destroyed
by the Pauli depairing field $H_P$, which can be estimated as $\mu_0
H_P\sim \Delta/\mu_B$, where $\mu_B$ is the Bohr magneton
\cite{Clogston62}. Comparing the two estimates we find $H_P/H_{c2}\sim
k_F\ell$, which shows that in materials which are close to the
metal-insulator transition, the Zeeman and orbital couplings are of
the same order of magnitude. This suggests that the suppression of
${\bar \Delta}$ with magnetic field in such samples should be
described qualitatively correctly by keeping only the Zeeman coupling,
of course only at sufficiently large $\Gamma$, where the transition is
continuous. This approximation has been used in the main text in
Figs.~3,4.

However, since the Zeeman and orbital couplings are of comparable
magnitude, it is legitimate to ask whether it is sufficient to keep
only the Zeeman coupling in calculating the effect of the magnetic
field on the density of states $N_b(\omega)$. To answer this question,
let us remember that, in a wide field range, type-II superconductors
exhibit the vortex state. If the density of states is to be measured
sufficiently far away from the vortex cores, as is assumed in this
work, then the orbital effect of the magnetic field can be taken into
account by the Doppler shift, which is proportional to the local
momentum of the supercurrent flow ${\bf q}$ in the point where the
density of states is being measured \cite{Schachinger00}. This changes
the bare electron Green's function in presence of magnetic field to
$$
\hat{G}_0^{-1}({\bf  k},\omega_n)=
(i\omega_n-b-\delta_{\bf k})\tau_0-\varepsilon_{\bf k}\tau_3,
$$
where $b$ is the Zeeman energy and 
$\delta_{\bf k}={\bf v}_{\bf k}\cdot{\bf q}$ is the Doppler shift.
Note that both pair-breaking fields $b$ and $\delta_{\bf k}$
enter the Green's function in the same way, the only difference
being that $\delta_{\bf k}$ depends on the direction of ${\bf k}$,
while $b$ is direction-independent. In presence of the Doppler shift,
the density of states changes to
\begin{equation}
N_b(\omega)=\frac{1}{4}\sum_{\pm}\int_0^\pi d\phi \sin\phi
N(\omega\pm b-v_Fq\cos \phi),
\label{eq:dos_doppler}
\end{equation}
which shows that the Doppler shift and the Zeeman coupling modify the
density of states in a similar fashion.

\begin{figure}[h]
\includegraphics[width=6.5cm]{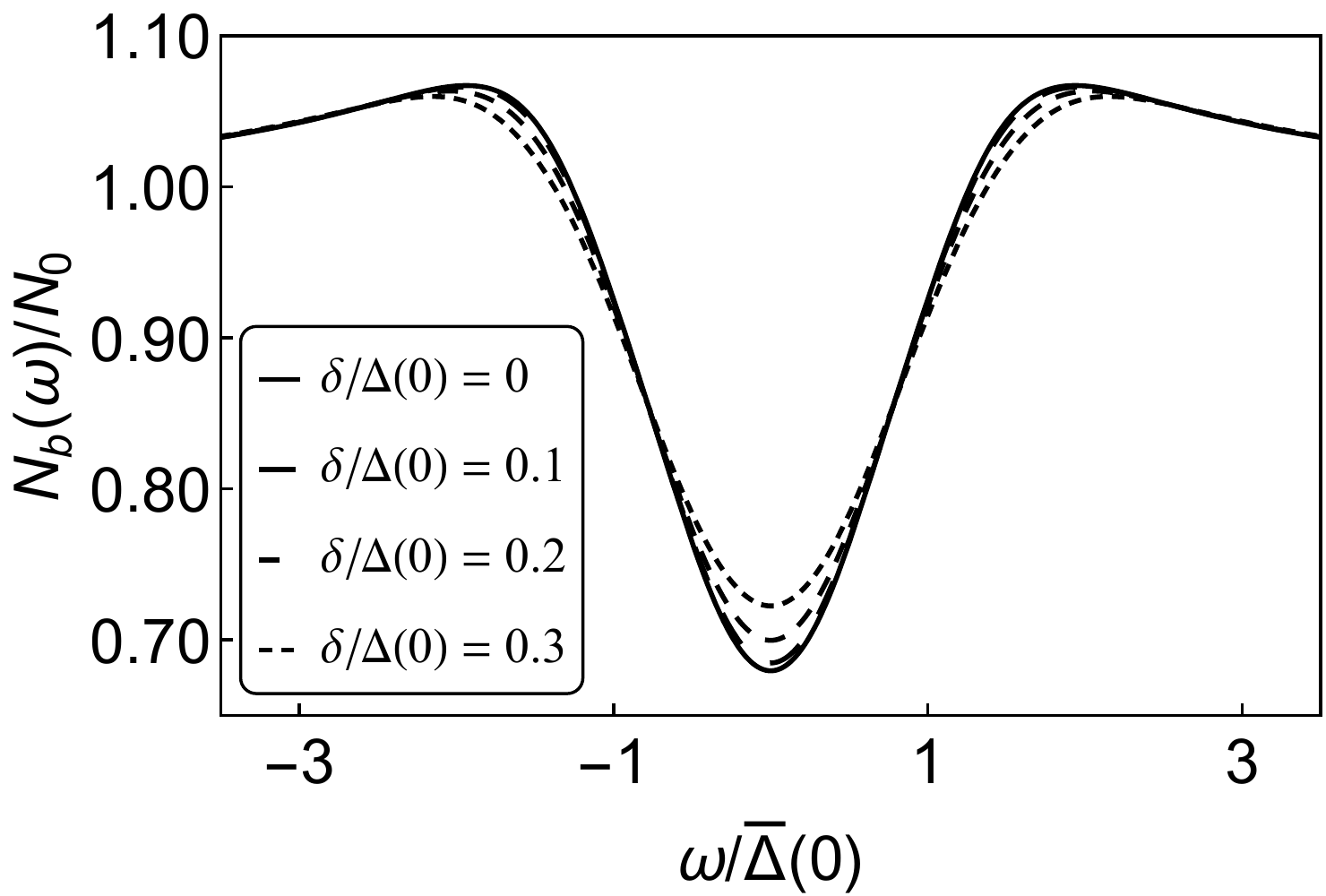}
\caption{Density of states $N_b(\omega)$ of a Dynes superconductor
  with $\Gamma/\Delta(0) = 0.38$ and Zeeman coupling $b/\Delta(0) =
  0.16$, when the gap is reduced to $\bar{\Delta}_b(0)/\Delta(0) =
  0.46$ (see Fig.~4 of the main text).  Note that the effect of the
  orbital coupling $\delta$ is very mild up to large values
  $\delta\sim 2b$. Moreover, the peak-to-peak distance of
  $N_b(\omega)$ exhibits further increase due to orbital effects.}
\label{fig:suppl_orbital}
\end{figure}

Finally, we need to fix the magnitude of $\delta=v_Fq$. Obviously,
$\delta$ is position-dependent, but it is easy to see that on the
boundaries of the flux-lattice cells, $\delta$ has to vanish by
symmetry. This means that the results presented in Fig.~4 of the main
text are directly applicable at such boundaries \cite{note_lattice}.
Moreover, Fig.~\ref{fig:suppl_orbital} shows that the orbital effects
on $N_b(\omega)$ are small with respect to the effect of the Zeeman
coupling up to large values of $\delta$, which shows that keeping only
the Zeeman coupling in estimating $N_b(\omega)$ should be a good
approximation in a quite broad range of positions away from the vortex
centers.

\section{Remarks on the experiment of Szab\'{o} et al.}
The differential tunneling conductance at a finite voltage $V$ between
a featureless normal metal and a superconductor with density of states
$N(\omega)$ is at finite temperatures given by
\begin{equation}
G(V)\propto \int N(\omega+eV)
\left(-\frac{\partial f}{\partial \omega}\right),
\label{eq:conductance}
\end{equation}
where $f(\omega)$ is the Fermi-Dirac distribution. Note that in the
zero-temperature limit $-\partial f/\partial \omega$ reduces to a
delta-function and $G(V)$ becomes directly proportional to $N(eV)$.
The fits shown in Fig.~1 of the main text were done making use of
Eq.~(\ref{eq:conductance}) with $f(\omega)$ taken at the finite
experimental temperature, and the density of states $N(\omega)$ was
described by the Dynes formula. Fitting parameters $\bar\Delta$ and
$\Gamma$ which have been used in those fits are shown in Table~I.

\begin{table}[h]
  \begin{tabular}{ | r | c | c | c | c |}
    \hline
    d(nm) & 3 & 5 & 10 & 30\\ \hline
    $\overline{\Delta}$(meV) & 0.19 & 0.63 & 1.12 & 1.22\\ \hline
    $\Gamma$(meV) & 0.16 & 0.21 & 0.1 & $10^{-3}$\\
    \hline
  \end{tabular}
\caption{Fitting parameters ${\bar\Delta}$ and $\Gamma$ which
  have been used in Fig.~1 of the main text for films with varying
  thickness $d$.}
\end{table}

Note that with decreasing film thickness $d$, the pair-breaking
parameter $\Gamma$ increases (the slight non-monotonicity of the
$\Gamma(d)$ dependence will be discussed later), while the ideal
superconducting gap ${\bar\Delta}$ decreases. Let us first discuss the
$d$-dependence of $\Gamma$. If our interpretation of the Dynes formula
in terms of the Lorentzian distribution of pair-breaking fields is
applicable to the data of Szab\'{o} et al. \cite{Szabo16}, then the
width of the distribution $P_m(V)$ has to increase with decreasing
$d$. This will obviously happen if the effective concentration of the
pair breakers grows with decreasing $d$. One possible scenario of how
this could happen is to assume that the pair breakers are located in
the vicinity of the interface between the film and the substrate.

\begin{figure}[h]
\includegraphics[width=6.5cm]{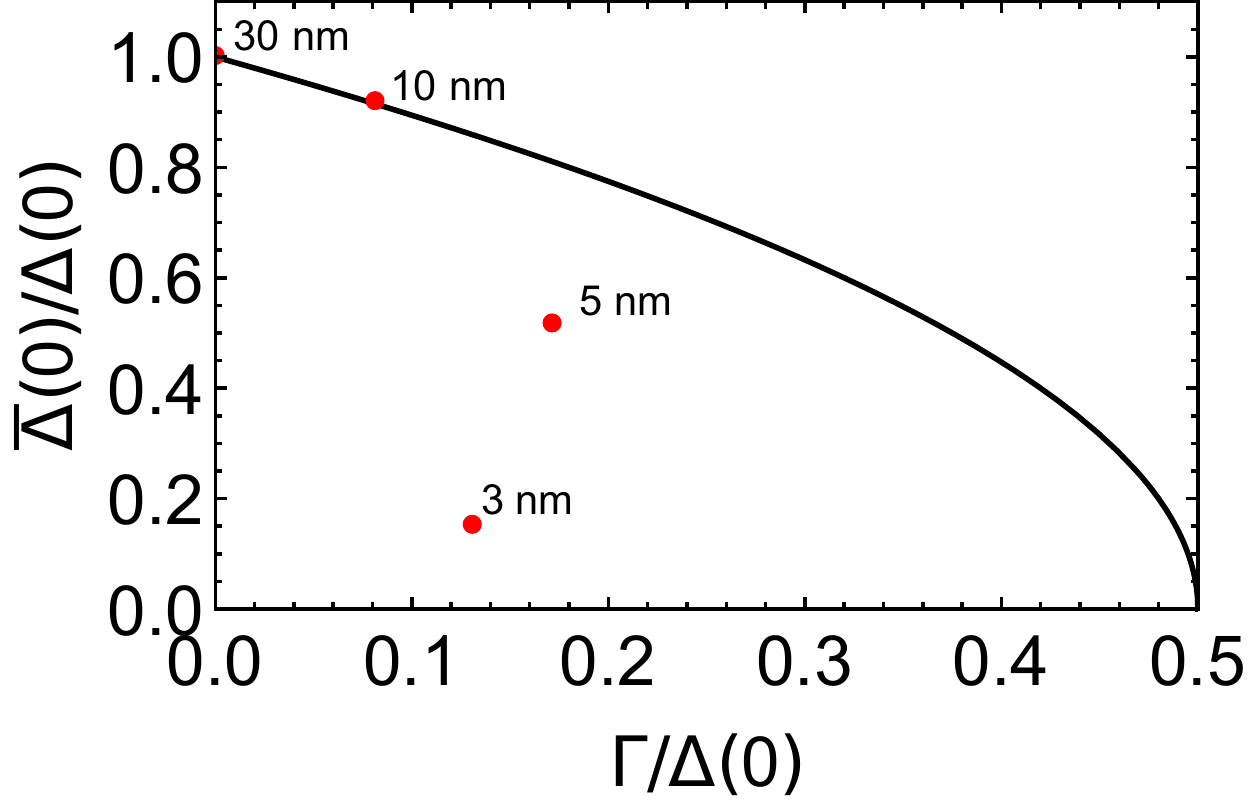}
\caption{Theoretical prediction for the evolution of the ideal
  superconducting gap ${\bar\Delta}(0)$ with the pair-breaking
  parameter $\Gamma$ of a Dynes superconductor. Experimental data are
  shown as red dots.}
\label{fig:suppl_szabo}
\end{figure}

Next we discuss the thickness dependence of ${\bar\Delta}$. Since
$\Gamma$ in the thickest sample is negligible and since $T\ll
{\bar\Delta}$, we will assume that the $T=0$ gap of a system without
pair breakers, $\Delta(0)$, is equal to the value of ${\bar\Delta}$
for $d=30$~nm, in other words $\Delta(0)=1.22$~meV. Switching on a
finite pair-breaking $\Gamma$ should lead then to a decrease of
${\bar\Delta}(0)$ described by
${\bar\Delta}(0)=\sqrt{\Delta(0)[\Delta(0)-2\Gamma]}$, see main text.
This prediction is shown in Fig.~\ref{fig:suppl_szabo}, together with
the experimental data taken from Table~I. Here we have assumed that
the $T=0$ values ${\bar\Delta}(0)$ can be approximated by the measured
values of ${\bar\Delta}$. This should be a good approximation, except
perhaps for the thinnest sample, whose $T_c$ is roughly only two times
larger than the experimental temperature. 

Figure~\ref{fig:suppl_szabo} shows that the initial decrease of
${\bar\Delta}(0)$ with increasing $\Gamma$ is captured well by our
theory. However, the agreement between theory and experiment breaks
down for the two thinnest films. This signals that different physical
phenomena, not included in our theory, start to play role in such very
thin films. We have learned recently that there are indications that
in those films, which are close to the Ioffe-Regel limit, the
normal-state density of states might exhibit the Altshuler-Aronov
singularity \cite{Grajcar16}. If this were true, then the normal-state
conductance $G_N(V)$ would not be constant and, in the most naive
approach, different $G(V)/G_N(V)$ curves would have to be fitted by
the Dynes formula. It is plausible that also the non-monotonic
behavior of $\Gamma(d)$ might be caused by the same physics.